\documentclass[12pt,english,draftcls, one column]{IEEEtran}
\usepackage[T1]{fontenc}
\usepackage[latin9]{inputenc}
\usepackage{color}
\usepackage{amsmath}
\usepackage{amssymb}
\usepackage{graphicx}
\usepackage{esint}

\makeatletter
\ifCLASSINFOpdf
\else
\fi
\usepackage{babel}

\makeatother

\usepackage{babel}
\begin{document}

\title{Modulation Classification via Gibbs Sampling Based on a Latent Dirichlet
Bayesian Network}

\author{Yu Liu, Osvaldo Simeone,\IEEEmembership{\ Member, IEEE,} Alexander
M. Haimovich,~\IEEEmembership{Fellow, IEEE,} Wei Su,~\IEEEmembership{Fellow, IEEE}
\thanks{Y. Liu, O. Simeone and A. M. Haimovich are with the Center for Wireless
Communications and Signal Processing Research (CWCSPR), ECE Department,
New Jersey Institute of Technology (NJIT), Newark, NJ 07102, USA (email:
\{yl227, osvaldo.simeone, haimovic\}@njit.edu). 

W. Su is with the U.S. Army Communication-Electronics Research Development
and Engineering Center, I2WD, Aberdeen Proving Ground, MD 21005, USA
(email: wei.su@ieee.org).%
}}
\maketitle
\begin{abstract}
A novel Bayesian modulation classification scheme is proposed for
a single-antenna system over frequency-selective fading channels.
The method is based on Gibbs sampling as applied to a latent Dirichlet
Bayesian network (BN). The use of the proposed latent Dirichlet BN
provides a systematic solution to the convergence problem encountered
by the conventional Gibbs sampling approach for modulation classification.
The method generalizes, and is shown to improve upon, the state of
the art. \end{abstract}

\section{Introduction}

\IEEEPARstart{R}{ecognition} of the modulation format of unknown
received signals is an important functionality of intelligent, or
cognitive, radios for both military and civilian applications \cite{Dobre}.
In most practical scenarios, the modulation classification task must
cope with the fact that transmission takes place over an unknown frequency-selective
channel. A generalized likelihood ratio test is proposed in \cite{N.lay}
to tackle this problem. The method, however, fails to identify nested
signal constellations such as \textcolor{black}{QPSK and 8-PSK} \cite{Panagiotou},
\cite{Hameed}. An alternative approach is to use a preliminary blind
equalization stage in order to compensate for the frequency-selective
channel \cite{Dobre}, \cite{Thomas}. The joint estimation of channel
and modulation is analytically intractable, and hence Markov Chain
Monte Carlo (MCMC) methods \cite{Doucet} provide viable solutions.
A classification method based on MCMC was proposed in \cite{Thomas}
for a single-antenna system by leveraging Gibbs sampling and by introducing
a ``superconstellation'' in order to solve the convergence problem
of conventional Gibbs sampling (see Sec. \ref{sec:Preliminaries}). 

In this paper, a novel Bayesian modulation classification scheme is
proposed for a single-antenna system over frequency-selective fading
channels. The method is based on Gibbs sampling as applied to a latent
Dirichlet Bayesian network (BN). The use of the proposed latent Dirichlet
BN provides a systematic solution to the convergence problem encountered
by the conventional Gibbs sampling approach. The method in \cite{Thomas}
based on ``superconstellation'' can be interpreted as an approximation
of the proposed approach. Furthermore, with an appropriate selection
of the prior distribution of the latent variable, our technique is
shown to improve over \cite{Thomas}.

\textit{Notation}: The superscripts \textit{T} and \textit{H} are
used to denote matrix or vector transpose and Hermitian, respectively.
We use the notation $\propto$ to denote proportionality. The cardinality
of a set $\mathcal{S}$ is denoted as $\left|\mathcal{S}\right|$.
Notation $\mathrm{\mathbf{1}}\left(\cdot\right)$ represents the indicator
function. The following notations for some important distributions
will be used: we denote by $\mathcal{CN}(\mathbf{\boldsymbol{\mu}},\mathbf{C})$
the circularly symmetric complex Gaussian distribution with mean vector
$\boldsymbol{\mu}$ and covariance matrix $\mathbf{C}$; we use $\mathcal{U}(\mathcal{S})$
to denote a uniform distribution on a set $\mathcal{S}$, i.e., all
elements of $S$ are equiprobable; notation $\mathrm{Cat}{}_{\mathcal{S}}\left(\mathbf{p}\right)$
is used for a categorical distribution on a set $\mathcal{S}$ with
a length-$\left|\mathcal{S}\right|$ vector $\mathbf{p}$ of probabilities;
the inverse gamma distribution with shape parameter $a$ and scale
parameter $b$ is denoted as $\mathcal{IG}\left(a,b\right)$.

\section{System Model}

We consider a single-antenna system over a frequency-selective fading
channel. The transmitted symbols $s_{n}$ are taken from a finite
constellation $A$, such as $M$-PSK or $M$-QAM. We assume that the
constellation $A$ belongs to a known finite set $\mathcal{A}$ of
possible constellations. The received baseband signal $r\left(t\right)$
at the output of the matched filter is given by
\begin{equation}
r\left(t\right)=\sum_{n=-\infty}^{\infty}s_{n}h\left(t-nT\right)+w(t),\label{eq:original system}
\end{equation}
where $h\left(t\right)$ represents the overall pulse shape, which
includes the effect of transmit and receive filters and of the channel
\cite{Barry}, and $w\left(t\right)$ is complex white Gaussian noise
with zero mean and variance $\sigma^{2}$. Without loss of generality,
the average power of the transmitted symbols is assumed to be unity,
i.e., $E(\left|s_{n}\right|^{2})=1$. Moreover, the pulse shape $h\left(t\right)$
is assumed to be of finite duration $LT$, for a given integer $L$.
Sampling at multiples of the symbol period $T$, the $k$th received
baseband sample is
\begin{equation}
r\left(kT\right)=\sum_{n=-\infty}^{\infty}s_{n}h\left(kT-nT\right)+w(kT).\label{eq: integer samples}
\end{equation}
Processing is performed in blocks of size $N$. Within this block,
the received samples (\ref{eq: integer samples}) can be collected
in an $N\times1$ vector $\mathbf{r}=[r(0),r(T),\cdots,r(\left(N-1\right)T)]^{T},$
which, using (\ref{eq: integer samples}), can be written as
\begin{equation}
\mathbf{r}=\mathbf{Sh\mathbf{\textrm{+}w}}.\label{eq:matrix rep}
\end{equation}
In (\ref{eq:matrix rep}), we have defined the $L\times1$ channel
vector $\mathbf{h}$ as $\mathbf{h}=[h(0),h(T),\cdots,h(\left(L-1\right)T)]^{T}$;
the vector $\mathbf{w}=[w(0),w(T),\cdots,w(\left(N-1\right)T)]^{T}\sim\mathcal{CN}(\mathbf{0},\sigma^{2}\mathbf{I})$
is the complex white Gaussian noise; and the $N\times L$ transmitted
symbol convolution matrix is defined as
\begin{equation}
\mathbf{S}=\left[\mathbf{s}_{0},\mathbf{s}_{1},\cdots,\mathbf{s}_{N-1}\right]^{T},
\end{equation}
where $\mathbf{s}_{n}=[s_{n},s_{n-1},\cdots,s_{n-L+1}]^{T}.$

Given the received signal vector $\mathbf{r}$ in (\ref{eq:matrix rep}),
the goal of the receiver is to classify the modulation format $A$
while being uninformed about the transmitted symbols $\mathbf{s}$,
defined as $\mathbf{s}=\left\{ s_{n}\right\} _{n=-L+1}^{N-1}$, the
channel vector $\mathbf{h}$ and the noise power $\sigma^{2}$.

\section{Preliminaries\label{sec:Preliminaries}}

In this work, as in \cite{Thomas}, we perform the modulation classification
task outlined above by using a Bayesian approach via MCMC methods.
In this section, we review some key preliminary concepts.

\subsection{Bayesian Approach}

The joint posterior probability density function (pdf) of the unknown
quantities $(A,\mathbf{s},\mathbf{\mathrm{\mathrm{\mathbf{h}},}\sigma}^{2})$
can be expressed as
\begin{equation}
p\left(A,\mathbf{s},\mathbf{h\mathrm{,\sigma^{2}}}\Big|\mathbf{r}\right)\propto p\left(\mathbf{r}\Big|A,\mathbf{s},\mathbf{h\mathrm{,\sigma^{2}}}\right)p\left(A,\mathbf{s},\mathbf{h},\mathbf{\mathrm{\sigma^{2}}}\right),\label{eq:posterior}
\end{equation}
where the likelihood function $p(\mathbf{r}|A,\mathbf{s},\mathbf{h\mathrm{,\sigma^{2}}})$
is such that
\begin{equation}
\mathbf{r}\Big|\left(A,\mathbf{s},\mathbf{h\mathrm{,\sigma^{2}}}\right)\sim\mathcal{CN}(\mathbf{Sh},\sigma^{2}\mathbf{I}),\label{eq:likelihood r}
\end{equation}
and the term $p(A,\mathbf{s},\mathbf{h},\mathbf{\mathrm{\sigma^{2}}})$
represents the available prior information on the unknown variables.
We assume that this prior distribution factorizes as 
\begin{equation}
p(A,\mathbf{s},\mathbf{h},\mathbf{\mathrm{\sigma^{2}}})=p\left(A\right)\left\{ \prod_{n=-L+1}^{N-1}p\left(s_{n}|A\right)\right\} p\left(\mathbf{h}\right)p\left(\sigma^{2}\right),\label{eq:7priors}
\end{equation}
where we have $A\sim\mathcal{U}\left(\mathcal{A}\right)$, $s_{n}|A\sim\mathcal{U}(A)$,
$\mathbf{h}\sim\mathcal{CN}(\mathbf{0},\alpha\mathbf{I})$ and $\sigma^{2}\sim\mathcal{IG}\left(\alpha_{0},\beta_{0}\right),$
where $(\alpha,\alpha_{0},\beta_{0})$ are fixed parameters. One typically
selects $\alpha$ and $\beta_{0}$ to be sufficiently large and $\alpha_{0}$
to be sufficiently small in order to obtain uninformative priors \cite{Thomas}. 

The Bayesian approach to modulation classification aims at estimating
the posterior probability of the modulation $A$ when conditioned
on the received signal $\mathbf{r}$, namely
\begin{equation}
p\left(A|\mathbf{r}\right)=\sum_{\mathbf{s}}\int p\left(A,\mathbf{s},\mathbf{h\mathrm{,\sigma^{2}}}|\mathbf{r}\right)d\mathbf{h}d\sigma^{2}.\label{eq:original Bayesian query}
\end{equation}
The computation of (\ref{eq:original Bayesian query}) involves a
multidimensional integration, which is generally infeasible. In the
following, we illustrate how this task can be accomplished by MCMC
techniques.

\subsection{Bayesian Network }

In order to facilitate the introduction of MCMC methods for modulation
classification in the next section, we first recall some basic facts
about BNs \cite{Koller}. A BN is a directed graph, whose nodes are
the random variables in the domain of interest and whose edges encode
the direct probabilistic influence of one variable on another. Specifically,
for a set of random variables $\left\{ X_{k}\right\} _{k=1}^{K}$,
a BN encodes a factorization of the joint distribution of the variables
at hand of the form

\begin{equation}
p\left(X_{1},\cdots,X_{K}\right)=\prod_{k=1}^{K}p\left(X_{k}|\mathrm{P}\mathrm{a}_{X_{k}}\right),\label{eq:BN' chain rule}
\end{equation}
where $\mathrm{P}\mathrm{a}_{X_{k}}$ represents a subset of the variables
$\left(X_{1},\cdots,X_{k-1}\right)$. By the chain rule, the factorization
(\ref{eq:BN' chain rule}) states that, when conditioning on all the
variables $\left(X_{1},\cdots,X_{k-1}\right)$, each variable $X_{k}$
is only influenced by the ``parent'' variables $\mathrm{P}\mathrm{a}_{X_{k}}$
(i.e., we have the Markov chain $\left(X_{1},\cdots,X_{k-1}\right)-\mathrm{P}\mathrm{a}_{X_{k}}-X_{k}$).
This statistical dependence between variable $X_{k}$ and the set
of parent variables $\mathrm{P}\mathrm{a}_{X_{k}}$ is encoded in
the BN by introducing a directed edge between all variables $\mathrm{P}\mathrm{a}_{X_{k}}$
and $X_{k}$. As an example, the BN encoding factorization (\ref{eq:posterior}),
(\ref{eq:7priors}) is shown in Fig. \ref{fig:BN-for-modulation}.

\subsection{Markov Chain Monte Carlo }

MCMC methods provide a general approach for generating $M$ samples
$\mathbf{x}^{\left(1\right)},\cdots,\mathbf{x}^{\left(M\right)}$
from an arbitrary target distribution $p\left(\mathbf{X}\right)$
with the aim of estimating ensemble averages, and hence multidimensional
integrals such as in (\ref{eq:original Bayesian query}). MCMC methods
simulate a Markov chain whose equilibrium distribution is $p\left(\mathbf{X}\right)$
in order to produce such samples \cite{Koller}. For instance, the
marginal of a joint distribution $p\left(\mathbf{X}\right)$ with
respect to any variable $X_{i}$ in $\mathbf{X}$ can be estimated
by MCMC as 

\begin{equation}
p\left(X_{i}=x_{i}\right)\approx\frac{1}{M}\sum_{m=M_{0}+1}^{M_{0}+M}\mathrm{\mathbf{1}}\left(\mathbf{x}_{i}^{\left(m\right)}=x_{i}\right),\label{eq:marginalization}
\end{equation}
where $\mathbf{x}_{i}^{\left(m\right)}$ is the $i$th element of
the $m$th sample of the simulated Markov chain and $x_{i}$ is a
value in the domain of $X_{i}$. Note that, in (\ref{eq:marginalization}),
the first $M_{0}$ samples generated by the Markov chain are not used
in order to limit the impact of the initialization. 

Gibbs sampling \cite{Koller} is a classical MCMC algorithm, whereby,
at each step $m$, a new sample of a given random variable $X_{i}$
is generated according to the conditional distribution $p(X_{i}|\mathbf{X}_{-i}=\mathbf{x}_{-i}^{\left(m-1\right)})$,
where $\mathbf{X}_{-i}$ denotes all variables in $\mathbf{X}$ except
$X_{i}$, which are fixed to the current value $\mathbf{x}_{-i}^{\left(m-1\right)}$.
Gibbs sampling is known to provide asymptotically correct estimates
(\ref{eq:marginalization}) (with probability one) under appropriate
conditions. A sufficient condition for convergence is that the conditional
distributions $p(X_{i}|\mathbf{X}_{-i})$ are strictly positive in
their domains for all $i$ \cite[Ch. 12]{Koller}. As we will see,
this condition is not satisfied by the distribution (\ref{eq:posterior})
for the problem under study. This suggests that more sophisticated
strategies than conventional Gibbs sampling are needed, as discussed
in the next section.

\section{Gibbs Sampling for Modulation Classification\label{sec:Bayesian-Modulation-Classificati}}

In this section, we design a Gibbs sampler that performs modulation
classification in the presence of frequency-selective fading. As outlined
in the previous section, the goal is to estimate the posterior probability
$p\left(A|\mathbf{r}\right)$ in (\ref{eq:original Bayesian query}).

\subsection{Conventional Gibbs Sampling }

\begin{figure}[htbp]
\begin{centering}
\textsf{\includegraphics[width=7.5cm,height=4cm]{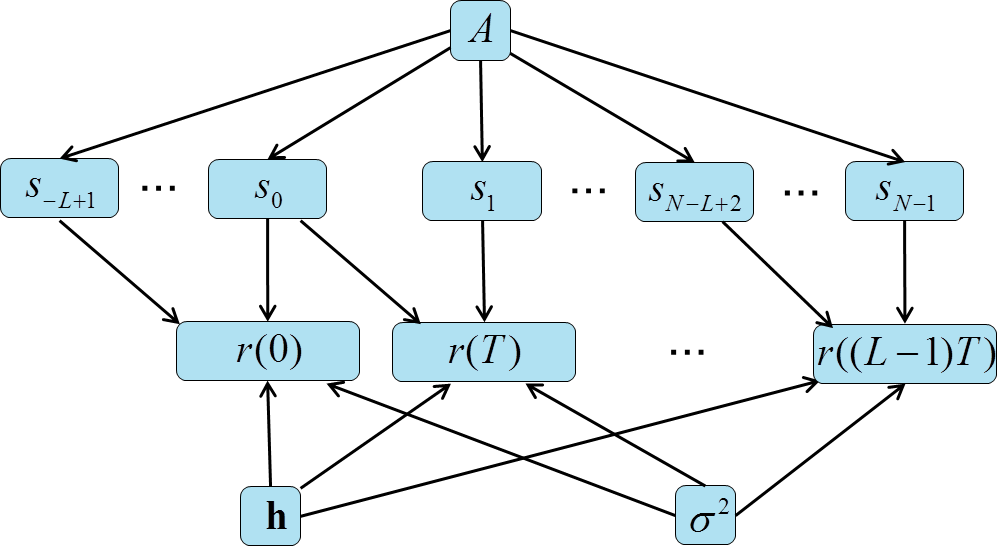}}
\par\end{centering}

\caption{\label{fig:BN-for-modulation}BN \textsl{$\mathcal{G}_{1}$} for modulation
classification based on the factorization (\ref{eq:posterior}).}
\end{figure}

We first elaborate on the conventional Gibbs sampler for the calculation
of the posterior $p\left(A|\mathbf{r}\right)$ that is based directly
on the joint distribution (\ref{eq:posterior})-(\ref{eq:7priors}).
The corresponding BN $\mathcal{G}_{1}$ is shown in Fig. \ref{fig:BN-for-modulation}.

As per the discussion in the previous section, Gibbs sampling requires
the knowledge of the conditional distributions of each variable given
all others. Calculating these conditional probabilities requires to
multiply all the factors in the factorization (\ref{eq:posterior})-(\ref{eq:7priors})
that contain the variable of interest and then normalize the resulting
distribution \cite[Ch. 12]{Koller} . This leads to the following
(see also \cite{Thomas}):
\begin{equation}
p\left(A\Big|\mathbf{s},\mathbf{\mathbf{\mathbf{h\mathrm{,\sigma^{2}}},\mathbf{r}}}\right)\propto p\left(A\right)\left\{ \prod_{n=-L+1}^{N-1}p\left(s_{n}|A\right)\right\} ,\label{eq:conditonal A}
\end{equation}
\begin{equation}
p\left(s_{n}\Big|A,\mathbf{s}_{-n},\mathbf{h\mathrm{,\sigma^{2},}}\mathbf{r}\right)\propto p\left(s_{n}|A\right)p\left(\mathbf{r}\Big|A,\mathbf{s},\mathbf{h\mathrm{,\sigma^{2}}}\right),\label{eq:conditional sn}
\end{equation}
\begin{equation}
\mathbf{h}\Big|\left(A,\mathbf{s},\mathbf{h\mathrm{,\sigma^{2},}}\mathbf{r}\right)\sim\mathcal{CN}(\mathbf{h}_{*},\boldsymbol{\Sigma}_{*}),\label{eq:conditional h}
\end{equation}
\begin{equation}
\mathrm{and}\,\,\sigma^{2}\Big|\left(A,\left\{ s_{n}\right\} _{n=-L+1}^{N-1},\mathbf{\mathbf{\mathbf{h\mathrm{,}}\mathbf{r}}}\right)\sim\mathcal{IG}\left(\alpha,\beta\right),\label{eq:contional sigma}
\end{equation}
where $h_{*}=\boldsymbol{\Sigma}_{*}(\mathbf{\boldsymbol{\Sigma}}_{0}^{-1}\mathbf{h}_{0}+\frac{\mathbf{S}^{H}\mathbf{r}}{\sigma^{2}}),$
$\mathbf{\boldsymbol{\Sigma}}_{*}^{-1}=\mathbf{\boldsymbol{\Sigma}}_{0}^{-1}+\frac{\mathbf{S}^{H}\mathbf{S}}{\sigma^{2}},$
$\alpha=\alpha_{0}+N$ and $\beta=\beta_{0}+\left\Vert \mathbf{r}-\mathbf{Sh}\right\Vert ^{2}$.
Note that (\ref{eq:conditional h}) follows from standard MMSE estimation
results (see, e.g., \cite{Koller}) and that (\ref{eq:contional sigma})
is a consequence of the fact that the inverse Gamma distribution is
the conjugate prior for the Gaussian likelihood \cite{Mruphy}. 

Gibbs sampling starts with an arbitrary feasible initialization for
all variables. In particular, one needs to initialize the constellation
$A$ to some value $A=a$ and, correspondingly, the transmitted symbols
$\mathbf{s}$ to source values belonging to the constellation $a$.
Using (\ref{eq:conditonal A}) and (\ref{eq:conditional sn}), it
can be easily seen that conventional Gibbs sampling will never select
values of $A$ different from the initial value $a$. This is due
to the fact that the conditional distribution $p\left(s_{n}|A\right)$
gives zero probability to all values of $s_{n}$ not belonging to
$A$. As a result, Gibbs sampling fails to converge to the posterior
distribution (see also Sec. \ref{sec:Preliminaries}). Next, we demonstrate
how this problem can be solved by the proposed approach based on the
latent Dirichlet BN.

\subsection{Gibbs Sampling Based on Latent Dirichlet BN}

\begin{figure}[htbp]
\begin{centering}
\textsf{\includegraphics[width=7.5cm,height=4cm]{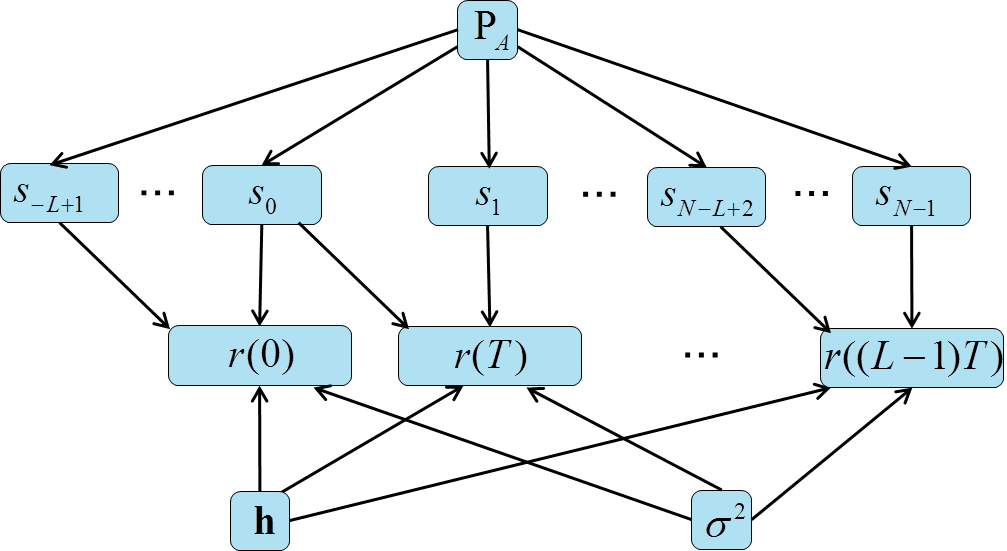}}
\par\end{centering}

\caption{\label{fig:New BN}BN \textsl{$\mathcal{G}_{2}$} for the proposed
modulation classification scheme based on the Dirichlet latent variable
$\mathbf{P}_{A}$.}
\end{figure}

In order to avoid the problem described above, we propose to base
the Gibbs sampler on the BN $\mathcal{G}_{2}$ shown in Fig. \ref{fig:New BN}.
In this BN, each transmitted symbol $s_{n}$ is distributed according
to a random mixture of uniform distributions on the different constellations.
Specifically, we introduce a random vector $\mathbf{P}_{A}$ to represent
the mixture weights, so that $\mathbf{P}_{A}\left(a\right)$ is the
probability that $s_{n}$ takes values in the constellation $a\mathcal{\in A}$.
The prior distribution of $\mathbf{P}_{A}$ is Dirichlet, so that
we have $\mathbf{P}_{A}\sim\mathrm{Dirichlet}\left(\boldsymbol{\gamma}\right)$
for a given set of nonnegative parameters $\boldsymbol{\gamma}=[\gamma_{1},\cdots,\gamma_{\left|\mathcal{A}\right|}]^{T}$
\cite{Koller}%
\footnote{Intuitively, the parameter $\gamma_{a}$ can be interpreted as the
number of symbols in constellation $a\in\mathcal{A}$ observed during
some preliminary measurements.%
}. When conditioned on $\mathbf{P}_{A}$, the transmitted symbol variables
$s_{n}$ are independent and distributed according to a mixture of
uniform distributions, i.e., $p\left(s_{n}|\mathbf{P}_{A}\right)=\sum_{a:\, s_{n}\in a}\mathbf{P}_{A}\left(a\right)/\left|a\right|$.

The BN $\mathcal{G}_{2}$, while departing from the original model
(\ref{eq:posterior})-(\ref{eq:7priors}), has the advantage that
a Gibbs sampler based on it is not limited by the zeros present in
the distribution (\ref{eq:posterior})-(\ref{eq:7priors}). In particular,
thanks to the introduction of the latent variable $\mathbf{P}_{A}$,
Gibbs sampling is able to explore different constellations irrespective
of its initialization. The idea of introducing the latent Dirichlet
variable $\mathbf{P}_{A}$ is inspired by \cite{Blei}, where a similar
quantity was used to account for the distribution of topics within
a document. According to the BN in Fig. \ref{fig:New BN}, the joint
pdf $p\left(\mathbf{P}_{A},\mathbf{s},\mathbf{h}\mathrm{,\sigma^{2}},\mathbf{r}\right)$
can be factorized as 
\begin{align}
 & p\left(\mathbf{P}_{A},\mathbf{s},\mathbf{h}\mathrm{,\sigma^{2}},\mathbf{r}\right)\nonumber \\
= & p\left(\mathbf{P}_{A}\right)\left\{ \prod_{n=-L+1}^{N-1}p\left(s_{n}|\mathbf{P}_{A}\right)\right\} p\left(\mathbf{h}\right)\cdot\nonumber \\
 & \cdot p\left(\sigma^{2}\right)p\left(\mathbf{r}\Big|\mathbf{s},\mathbf{h\mathrm{,\sigma^{2}}}\right),\label{eq: factorization 2}
\end{align}
where, as discussed above, we have $\mathbf{P}_{A}\sim\mathrm{Dirichlet}(\boldsymbol{\gamma})$
and $p\left(s_{n}|\mathbf{P}_{A}\right)=\sum_{a:\, s_{n}\in a}\mathbf{P}_{A}\left(a\right)/\left|a\right|$,
while the remaining conditional distributions are as in (\ref{eq:likelihood r})
and (\ref{eq:7priors}).

To apply Gibbs sampling based on the factorization (\ref{eq: factorization 2}),
the conditional distributions for $\mathbf{P}_{A}$, $s_{n}$, $\mathbf{h}$
and $\mathrm{\sigma^{2}}$ conditioned on all other variables are
required. It can be shown that the conditional distribution for $\mathbf{h}$
and $\sigma^{2}$ are (\ref{eq:conditional h}) and (\ref{eq:contional sigma})
respectively. The other required conditional distributions of $\mathbf{P}_{A}$
and $s_{n}$ are as follows:
\begin{equation}
\mathbf{P}_{A}\Big|\left(\mathbf{s},\mathbf{h}\mathrm{,\sigma^{2}},\mathbf{r}\right)\sim\mathrm{Dirichlet}\left(\boldsymbol{\gamma}+\mathbf{c}\right),\label{eq:conditonal Pa}
\end{equation}
 where $\mathbf{c}=\left[c_{1},\cdots,c_{\left|\mathcal{A}\right|}\right]^{T}$
and $c_{a}$ is the number of symbols that belong to constellation
$a\in\mathcal{A}$;
\begin{align}
\mathrm{and}\,\, & p\left(s_{n}\Big|\mathbf{P}_{A},\left\{ s_{k}\right\} _{k\neq n}^{N-1},\mathbf{h\mathrm{,\sigma^{2},}}\mathbf{r}\right)\nonumber \\
\propto & p\left(s_{n}|\mathbf{P}_{A}\right)p\left(\mathbf{r}\Big|\mathbf{s},\mathbf{h\mathrm{,\sigma^{2}}}\right).
\end{align}
Note that (\ref{eq:conditonal Pa}) follows the fact that different
modes of the mixture distribution is distributed as a categorical
distribution and Dirichlet distribution is the conjugate prior for
categorical distribution \cite{Koller}.

The task of modulation classification is achieved by computing the
posterior distribution $p\left(\mathbf{P}_{A}|\mathbf{r}\right)$
following the Gibbs procedure discussed above. From the posterior
$p\left(\mathbf{P}_{A}|\mathbf{r}\right)$, we can then obtain an
estimate $\hat{A}$ for the constellation as
\begin{equation}
\hat{A}=\arg\max_{a\in\mathcal{A}}E\left[\mathbf{P}_{A}\left(a\right)\mid\mathbf{r}\right],
\end{equation}
where the expectation is taken over the distribution $p\left(\mathbf{P}_{A}|\mathbf{r}\right)$.

\textit{Remark 1}: The method proposed in \cite{Thomas}, based on
the introduction of a ``superconstellation'', can be seen as an
approximation of the approach presented above. Specifically, the scheme
of \cite{Thomas} is obtained by setting $\boldsymbol{\gamma}=\mathbf{0}$
and by choosing $\mathbf{P}_{A}$ to be equal to $\mathbf{c}/\sum_{a=1}^{\left|\mathcal{A}\right|}c_{a}$,
at each Gibbs iteration, where we recall that $c_{a}$ is the number
of symbols that belong to constellation $a\in\mathcal{A}$. \textcolor{black}{Furthermore,
the computational complexity of the proposed scheme is comparable
to the superconstellation Gibbs sampler}\textcolor{blue}{{} }\textcolor{black}{\cite[Table I]{Thomas}. }

\textit{Remark 2}: At high signal-to-noise ratios (SNR), the relationship
between $\mathbf{r}$ and $\mathbf{s}$ defined by (\ref{eq:likelihood r})
is almost deterministic. Following the discussion in the previous
section, this may create convergence problems. This issue can be tackled
via the idea of annealing \cite[Ch. 12]{Koller}. Accordingly, the
distribution (\ref{eq:likelihood r}) is modified as $\mathbf{r}\Big|\left(A,\mathbf{s},\mathbf{h\mathrm{,\sigma^{2}}}\right)\sim\mathcal{CN}(\mathbf{Sh},\rho\sigma^{2}\mathbf{I})$,
where $\rho$ is a ``temperature'' parameter. The procedure starts
with a high value of $\rho$ to prevent the mentioned convergence
problems, and then cools to a lower temperature to produce the desired
target distribution. Effective cooling schedules include logarithmic
and linear decreases of the temperature, whose parameters can be determined
based on preliminary runs \cite{Nourani}.

\section{Numerical Results and Concluding Remarks}

In this section, we evaluate the performance of the proposed scheme
for the recognition of three modulation formats, namely QPSK, 8-PSK
and 16-QAM. We assume Rayleigh fading channels, which are normalized
so that $E[\left\Vert \mathbf{h}\right\Vert ^{2}]=1$. The average
SNR is defined as SNR$=1/\mathrm{\sigma^{2}}$. The number of samples
used by Gibbs sampling are $M=300$ and $M_{0}=100$ in (\ref{eq:marginalization}).
No annealing is used. The performance criterion of interest is probability
of correct classification (PCC).

\begin{figure}[htbp]
\begin{centering}
\textsf{\includegraphics[width=8.5cm,height=6.5cm]{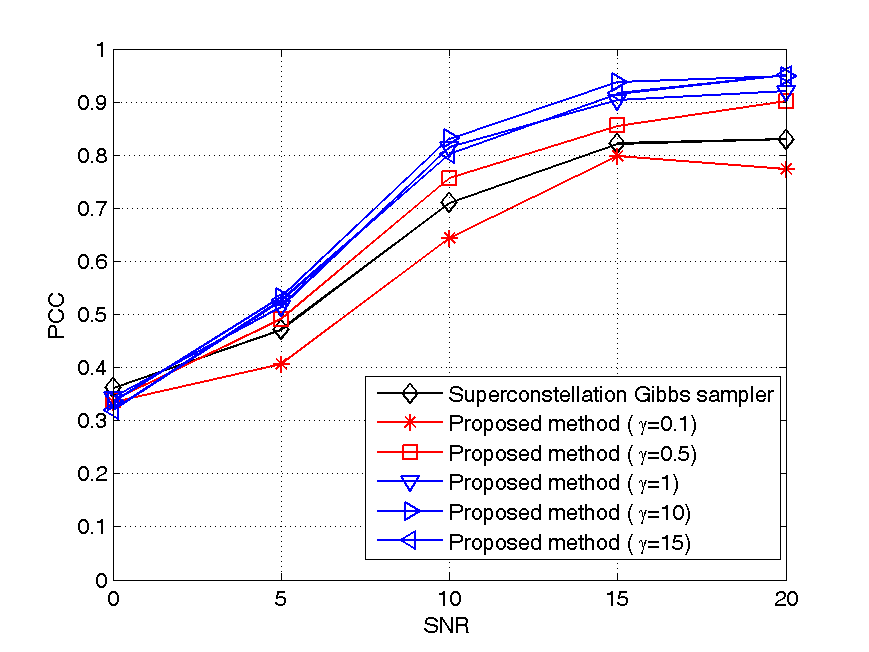}}
\par\end{centering}

\caption{\label{fig:PCC1}Probability of correct classification using the proposed
method and the superconstellation Gibbs sampling of \cite{Thomas}
versus SNR with $L=3$ independent taps.}
\end{figure}

In Fig. \ref{fig:PCC1}, we plot the PCC for $L=3$ independent taps
with relative powers given by $\left[0\mathrm{dB},-0.9\mathrm{dB},-4.9\mathrm{dB}\right]$.
The performance of the proposed method is compared to the superconstellation
Gibbs sampler of \cite{Thomas}. The prior distribution $\mathrm{Dirichlet}\left(\boldsymbol{\gamma}\right)$
for $\mathbf{P}_{A}$ is selected so that all elements of the vector
$\mathbf{\mathbf{\boldsymbol{\gamma}}}$ are identical and equal to
a parameter $\gamma$. In order to investigate the impact of prior
distributions for $\mathbf{P}_{A}$ on the classification performance,
five values, $0.1$, $0.5$, $1$, $10$ and $15$ are considered
for $\gamma$. \textcolor{black}{It is observed that the performance
is enhanced with a larger $\gamma$, especially for higher SNR values.
This is because increasing $\gamma$ enhances the relative importance
of the prior distribution and helps improve the convergence properties
of the algorithm (see Remark 2).} \textcolor{black}{We observe that
in practice the value of the hyperparameter $\gamma$, as is the case
for all of MCMC methods \cite{Neal}, can be determined based on offline
preliminary runs.} \textcolor{black}{Finally, for sufficiently high
SNRs, the superconstellation Gibbs sampler achieves a PCC of about
$83\%$, while the proposed scheme achieves a PCC of $95\%$ with
$\gamma=15$. }

\begin{figure}[htbp]
\begin{centering}
\textsf{\includegraphics[width=8.5cm,height=6.5cm]{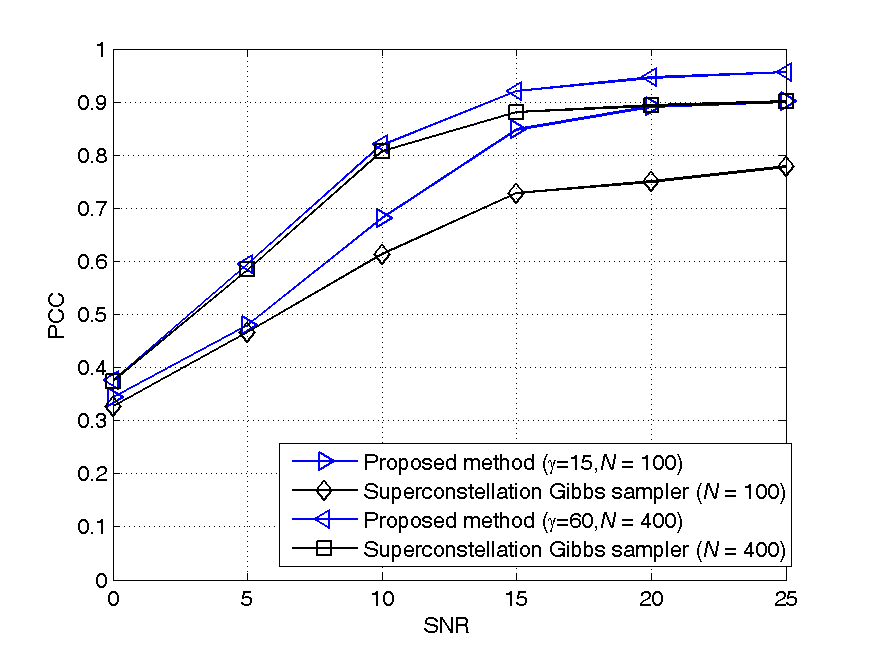}}
\par\end{centering}

\caption{\label{fig:PCC1-1}Probability of correct classification using the
proposed method and the superconstellation Gibbs sampling of \cite{Thomas}
versus SNR with two paths and non-integer delays ($L=6$).}
\end{figure}

Fig. \ref{fig:PCC1-1} shows the PCC for channels with two paths with
non-integer delays $\left[0,1.3T\right]$ and with relative powers
$\left[0\mathrm{dB},-0.9\mathrm{dB}\right]$ \cite[Ch. 3]{Goldsmith}.
A raised cosine pulse shape with roll-off factor 0.3 is assumed and
we set $L=6$. For $N=100$, while there is performance degradation
as compared to 3-tap channels due to more severe frequency selectivity,
the proposed scheme still can achieve above $90\%$ PCC at sufficiently
large SNR, while the method of \cite{Thomas} achieves a PCC of $80\%$.
\textcolor{black}{For $N=400$, both classification schemes obtain
performance gains, but the proposed scheme attains a PCC of $95.6\%$,
while the superconstellation method achieves a PCC of $90\%$.}

\textcolor{black}{In summary, as demonstrated by the discussed numerical
results, the proposed Gibbs sampling method based on latent Dirichlet
Bayesian network provides significant performance gains over the state
of the art.}


\begin{thebibliography}{10}
\bibitem{Dobre}O. A. Dobre, A. Abdi, Y. Bar-Ness and W. Su, \textquotedblleft{}A
survey of automatic modulation classification techniques: classical
approaches and new developments,\textquotedblright{} \textit{IET Communications},
vol.1, no. 2, pp. 137-156, Apr. 2007. 

\bibitem{N.lay}N. Lay and A. Polydoros, ``Modulation classification
of signals in unknown ISI environments,'' in \textit{Proc. IEEE MILCOM},
pp. 170-174, San Diego, CA, Nov. 1995.

\bibitem{Panagiotou}P. Panagiotou, A. Anastasopoulos, and A. Polydoros,
\textquotedblleft{}Likelihood ratio tests for modulation classification,\textquotedbl{}
in \textit{Proc. IEEE MILCOM}, pp. 670-674, Los Angeles, CA, Oct.
2000, 

\bibitem{Hameed}F. Hameed, O. A. Dobre, and D. C. Popescu, \textquotedblleft{}On
the likelihood-based approach to modulation classification,\textquotedbl{}
\textit{IEEE Trans. Wireless Commun}., vol. 8, no. 12, pp. 5884-5892,
Dec. 2009. 

\bibitem{Thomas}T. A. Drumright and Z. Ding. ``QAM constellation
classication based on statistical sampling for linear distortive channels,''
\textit{IEEE Trans. Signal Process}., vol. 54, no. 5, pp. 1575-1586,
May 2006.

\bibitem{Doucet}A. Doucet and X. Wang, \textquotedblleft{}Monte Carlo
methods for signal processing,\textquotedblright{} \textit{IEEE Signal
Process}. \textit{Mag}., vol. 22, no. 6, pp. 152\textendash{}170,
Nov. 2005.

\bibitem{Barry}J. R. Barry, E. A. Lee and D. G. Messerschmitt, \textit{Digital
Communication}, Springer, 2003.

\bibitem{Koller}D. Koller and N. Friedman, \textit{Probabilistic
Graphical Models: Principles and Techniques}, MIT Press, 2009. 

\bibitem{Mruphy}K. P. Murphy, \textquotedblleft{}Conjugate Bayesian
analysis of the Gaussian distribution,\textquotedblright{} \textit{Univ.
of British Columbia, Canada, Tech. Rep}., 2007 {[}Online{]}. Available:
http://www.cs.ubc.ca/\textasciitilde{}murphyk/Papers/bayesGauss.pdf

\bibitem{Blei}D. Blei, A. Ng, and M. Jordan. ``Latent Dirichlet
allocation,'' \textit{Journal of Machine Learning Research}, vol.
3, pp. 993\textendash{}1022, Jan. 2003. 

\bibitem{Nourani}Y. Nourani and B. Andresen, \textquotedblleft{}A
comparison of simulated annealing cooling strategies,\textquotedblright{}\textit{
J. Phys}. \textit{A}, vol. 31, no. 41, pp. 8373-8385, July 1998.

\bibitem{Neal}\textcolor{blue}{{} }\textcolor{black}{R. M. Neal, ``Sampling
from multimodal distributions using tempered transitions,'' }\textit{\textcolor{black}{Statistics
and computing}}\textcolor{black}{, vol. 6, no. 4, pp. 353-366, June
1996. }

\bibitem{Goldsmith}A. Goldsmith, \textit{Wireless Communications},
Cambridge University Press, 2005.\end{thebibliography}
\end{document}